# Reciprocal and Extortive Strategies: Infinitely Iterated Prisoner's Dilemma

Robert D. Young*

*Department of Physics, Illinois State University,
Normal, IL 61790-4560*

(Dated November 25, 2019)

**Abstract**

The Prisoner's Dilemma game has a long history stretching across the social, biological, and physical sciences. In 2012, Press and Dyson developed a method for analyzing the mapping of the 8-dimensional strategy profile onto the 2-dmensional payoff space in an infinitely iterated Prisoner's Dilemma game, based on Markov chain analysis with memory-one strategies. We generalize this approach and introduce the concept of strategy parameter to show that linear relations among player payoffs are a ubiquitous feature of the infinitely iterated Prisoner's Dilemma game. Our extended analysis is applied to various strategy profiles including tit-for-tat, win-stay-lose-shift, and other randomized strategy sets. Strategy profiles are identified that map onto the vertices, edges, and interior of the Prisoner's Dilemma quadrilateral in the two-dimensional payoff (score) space. A DaMD strategy is defined based solely on "Defection after Mutual Defection" and leads to linear relations between player scores using strategy parameter analysis. The DaMD strategy is shown to result in an equal (reciprocal) or larger (extortive) score for its user compared to the other player, independent of the strategy set of the other player. The extortive scores occur when the probabilities for the DaMD player to cooperate after conflicting plays (cooperate-defect or defect-cooperate) sum to less than 1. The equal reciprocal scores occur when the probabilities for the DaMD player to cooperate after conflicting plays (cooperate-defect or defect-cooperate) sum to 1. When one player selects the extortive DaMD, the opposing player can force the equal punishment payoffs for both players in the infinitely iterated Prisoner's Dilemma by also choosing the DaMD strategy. Possible pathways to mutual cooperation based on DaMD are discussed.

PACS Codes: 02.50.Ga; 02.50.Le; 89.65.-s; 89.75.Fb

## I. Introduction

Prisoner's Dilemma is a time-honored paradigm for 2X2 games that are used for understanding of complex problems in the social, behavioral, biological, and physical sciences [1,2,3,4]. Prisoner's Dilemma (PD) was developed in 1950 at RAND by Flood and Dresher [2]. Axelrod describes many of the strategies used to solve the iterated PD and reports on "tournaments" to test various strategies developed by himself and many others [3]. These tournaments have continued in one form or another [5]. Nowak has authored an accessible book which treats the iterated PD as well as other classic iterated games focusing on evolutionary dynamics [6]. Recently, Lambert, Vyawahare, and Austin have applied a game theory approach to the physics of bacteria growth and, potentially, cancer propagation [7]. The scope and long history of game theory, Prisoner's Dilemma, and applications makes it surprising that a novel analytic technique for infinitely iterated Prisoner's Dilemma (IPD) was discovered by Press and Dyson in 2012 [8]. At its core, this method of analysis provides tools for understanding the mapping of the 8-dimensional hypercube of strategy profiles of IPD onto the 2-dimensional space of payoffs for the 2 players. This discovery led to a resurgence of interest and new applications of the Prisoner's Dilemma, mostly in evolutionary dynamics [9,10,11,12]. The zero-determinant strategy (ZDS) assumes a player always defects after mutual defection along with other conditions on the player's actions [8]. ZDS was shown to result in a linear relationship between scores of the players. A ZDS that received a higher score than a fully cooperative opponent was called extortionate.

* Email: rdyoung@ilstu.edu

We develop a general analytic technique using the concept of strategy parameter. We identify strategy sets that lead to vertex scores of the PD stage game and strategy sets that map onto the edges of the stage game. These edge strategy sets correspond to players who fully defect or fully cooperate independent of the strategy set of the other player. We show, using a strategy parameter analysis, that linear relations between player scores are a ubiquitous feature of the infinitely iterated Prisoner's Dilemma for many well-known strategy sets. Strategy sets include tit-for-tat (TFT), win-stay-lose-shift (WSLS), randomized versions of TFT and WSLS, and a randomized strategy set called the $\varepsilon$-strategy set which is an extended version of Axelrod's RANDOM strategy set [3,5,6,9,11]. TFT and WSLS also play vs an opponent using an arbitrary strategy. The initial Axelrod tournaments for IPD were won by TFT [3]. Nowak now calls WSLS "currently 'world champion'" of play for the IPD [6]. The linear relations generated using our strategy parameter analysis give theoretical underpinnings for these computer tournament results.

Based on strategy parameter analysis, we define a general strategy set (DaMD, "Defection after Mutual Defection") that subsumes the ZDS and that can generate a score for its user that is larger than or equal to the score of the other player. We call these two classes of DaMD – extortive and reciprocal. This score ordering is independent of the opponent strategy unless the opponent also uses DaMD. We show that other strategies besides DaMD and ZDS can also result in the same type of behavior of scores, but with an important difference. WSLS is a particularly surprising strategy set in that it has extortive and reciprocal score ordering, but the outcomes are decided by the play of the opponent. Mutual DaMD strategies, leading to a string of mutual defection, are the equilibrium strategy profile for two rational players in an infinitely iterated Prisoner's Dilemma game. This is the same result as the equilibrium mutual defection strategy profile for the single-play and finitely iterated Prisoner's Dilemma games. The DaMD, however, describes a possible pathway to cooperation and enhanced scores for both players, including rational and naïve players. The result can be stable scores greater than those of mutual defection for both players.

## II. Prisoner's Dilemma Stage Game

The focus here is the Iterated Prisoner's Dilemma (IPD) game. The IPD is based on the static, stage game that is described in terms of a 2X2 matrix with ordered pair elements. This matrix captures both the strategies and the payoffs of the players in the stage game [13,14]. Fig. 1 gives a standard form of the matrix.

$$\begin{array}{c} \quad\quad\quad \mathbf{Y} \;\; c \quad\quad\;\; d \\ \mathbf{X} \begin{array}{c} c \\ d \end{array} \begin{bmatrix} (R,R') & (S,T') \\ (T,S') & (P,P') \end{bmatrix} \end{array}$$

Figure 1. Prisoner's Dilemma Matrix.
See text for discussion.

In Fig. 1, the two players are labeled X and Y. Decisions or actions for the individual players are labeled by c (cooperate) and d (defect). The ordered pairs in the four quadrants enclosed by brackets are payoffs of the two players. The first entry of the ordered pair corresponds to the payoff for player X, the second entry to the payoff for player Y. The analysis here is not restricted to having the payoffs for X and Y the same, but we shall assume this condition resulting in a symmetric stage game. So $R = R'$, etc. If both players cooperate, the payoff is $R$ (reward) to each. If both player defect, the payoff is $P$ (punishment) to each. If one player cooperates and the other player defects, then the player that cooperates gets payoff $S$ (sucker) and the player that defects get payoff $T$ (temptation). The classic PD stage game puts two conditions on the four payoffs: $T > R > P > S$ and, less crucial to PD, $2R > T + S$. The first condition ensures that mutual defection with payoff $(P, P)$ is a Nash equilibrium strategy [13,14,15]. The second condition ensures that mutual cooperation with payoff $(R, R)$ is the best mutual outcome of the four ordered pairs, a Pareto optimal payoff for both players [13,14]. There are no other conditions on the payoffs for the PD except that usually the probabilities and payoffs employed are drawn from rational numbers. Axelrod emphasizes the payoff set $(T, R, P, S) = (5, 3, 1, 0)$ which satisfies the condition $2R > T + S$ [3]. The general analysis below does not depend on the specific values of the payoff set. In all numerical examples, we use the traditional Axelrod payoffs although general results are valid for any payoff set that satisfies the two conditions above. The matrix in Fig. 1 includes both player payoffs (ordered pairs) and player actions (c or d). Game theory uses terms with social, behavioral, or economic connotations so such terms (game, player, payoff, strategy, cooperate, defect, decision, etc.) are encountered in various sections but are concentrated in Section IX. Conclusions.

The payoff sets for the static stage game are frequently depicted on a 2-dimensional plane in which the payoffs of the two players are denoted by $S_i = T, R, P,$ or $S$ with $i = X, Y$. Fig. 2 depicts the payoff sets using the Axelrod values.

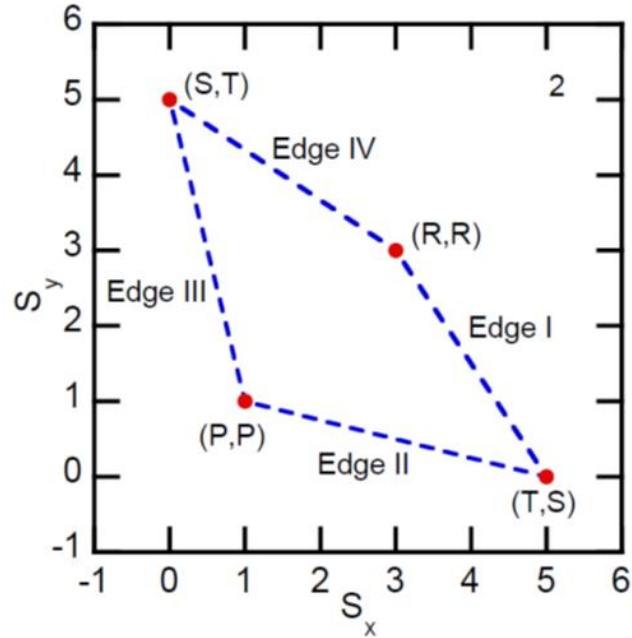

Figure 2. Prisoner's Dilemma stage game.
(T,R,P,S) = (5,3,1,0). See text for discussion.

For a static stage game with single play and synchronous actions for X and Y, only the four vertices shown as solid circular points in Fig. 2 are valid payoffs. However, it has been realized even from the original work of von Neumann and Nash, that some form of repeated game was necessary for game theory to be relevant to real conflicts. This realization in turn leads to the payoff sets within the quadrilateral enclosed by the dashed lines including the boundary as being valid payoffs for the IPD. The characterization of the strategy profiles for both players leading to these more general payoff points ($S_X, S_Y$) is addressed below. In preparation for this analysis, the equations for the straight lines forming the four edges of the quadrilateral in Fig. 2 are listed in Eq. 1 together with the equations for Axelrod values.

i) Edge I: $(R-S)S_X + (T-R)S_Y - R(T-S) = 0$ ; $3S_X + 2S_Y - 15 = 0$.

ii) Edge II: $(P-S)S_X + (T-P)S_Y - P(T-S) = 0$ ; $S_X + 4S_Y - 5 = 0$. (1)

iii) Edge III: $(T-P)S_X + (P-S)S_Y - P(T-S) = 0$ ; $4S_X + S_Y - 5 = 0$.

iv) Edge IV: $(T-R)S_X + (R-S)S_Y - R(T-S) = 0$ ; $2S_X + 3S_Y - 15 = 0$.

At this point, the linear relations in Eq. 1 are only algebraic formulas for the dashed lines joining the four vertices in Fig. 2. Below we show that the four edges have meaning for the IPD in terms of mapping of specific strategy profiles.

### III. Generalized Analysis of Iterated Prisoner's Dilemma

The iterated Prisoner's Dilemma involves infinitely repeated play of the PD stage game and synchronous actions of players X and Y. The assumptions of the IPD result in transition probability matrices for player actions that are Markov with row entries adding to unity. The theory of infinite Markov chains is the fundamental theory for IPD [16,17]. We review the essential concepts of the Markov chain approach with memory-one strategies to the iterated Prisoner's Dilemma [8]. The 4-vector $\mathbf{p}=(p_1,p_2,p_3,p_4)$ is the strategy set of player X. $p_i$ are conditional transition probabilities for player X to cooperate in the current round of the stage game, and $0 \le p_i \le 1$. The 4-vector $\mathbf{q}=(q_1,q_3,p_2,p_4)$ is the strategy set of Y. $q_i$ are also conditional transition probabilities for Y to cooperate so $0 \le q_i \le 1$. The probabilities are:
$p_1 = \text{Prob}(X \to c \mid cc)$; $p_2 = \text{Prob}(X \to c \mid cd)$; $p_3 = \text{Prob}(X \to c \mid dc)$; $p_4 = \text{Prob}(X \to c \mid dd)$
The probabilities for Y are defined similarly except that $q_2 = \text{Prob}(Y \to c \mid dc)$ and $q_3 = \text{Prob}(Y \to c \mid cd)$. Formally, all probabilities, payoffs and scores are rational numbers. Payoffs can be cardinal or ordinal. The IPD strategy profile $\{\mathbf{p};\mathbf{q}\}$ forms a unit hypercube in the 8-dimensional strategy space. The Markov transition matrix $\mathbf{M}$ is defined, for example, in Nowak (p. 83) [6]. $\mathbf{M}$ has a unit eigenvalue. The matrix $\overline{\mathbf{M}} \equiv \mathbf{M} - \mathbf{I}$ is singular with zero determinant and stationary row eigenvector $\mathbf{v}$ so $\mathbf{v}\overline{\mathbf{M}} = \mathbf{0}$. The scaler product of an arbitrary vector $\mathbf{f}$, given by the transpose $\mathbf{f}^T = (f_1,f_2,f_3,f_4)$, and the stationary probability vector $\mathbf{v}$ is given by $\mathbf{v} \cdot \mathbf{f} = \mu D(\mathbf{p},\mathbf{q},\mathbf{f})$ where $\mu$ is a constant. The average of $\mathbf{f}$ for the Markov chain states is then given by

$$\langle \mathbf{f} \rangle = \frac{\mathbf{v} \cdot \mathbf{f}}{\mathbf{v} \cdot \mathbf{1}} = \frac{\mu D(\mathbf{p},\mathbf{q},\mathbf{f})}{\mu D(\mathbf{p},\mathbf{q},\mathbf{1})} = \frac{D(\mathbf{p},\mathbf{q},\mathbf{f})}{D(\mathbf{p},\mathbf{q},\mathbf{1})}, \qquad (2)$$

where $\mathbf{1}^T = (1,1,1,1)$ and $\mathbf{v} \cdot \mathbf{1}$ is needed for normalization. The determinant $D(\mathbf{p},\mathbf{q},\mathbf{f})$ is

$$D(\mathbf{p},\mathbf{q},\mathbf{f}) = \det \begin{vmatrix} -1+p_1 q_1 & -1+p_1 & -1+q_1 & f_1 \\ p_2 q_3 & -1+p_2 & q_3 & f_2 \\ p_3 q_2 & p_3 & -1+q_2 & f_3 \\ p_4 q_4 & p_4 & q_4 & f_4 \end{vmatrix}, \qquad (3)$$

$D(\mathbf{p},\mathbf{q},\mathbf{f})$ is obtained from the matrix $\overline{\mathbf{M}}$ with the vector $\mathbf{f}$ using a standard set of matrix and determinant properties and manipulations as described eloquently by Press and Dyson [8]. Inspection of the determinant $D(\mathbf{p},\mathbf{q},\mathbf{f})$ shows that the second column depends on only the strategy $\mathbf{p}$ of player X while the third column depends on only the strategy $\mathbf{q}$ of player Y. The determinant $D(\mathbf{p},\mathbf{q},\mathbf{f})$ has a simplified, more intuitive structure compared to the Markov matrices $\mathbf{M}$ or $\overline{\mathbf{M}}$. The payoff vector for X in the stage game is $\mathbf{S}_X = (R,S,T,P)$ and that for Y, $\mathbf{S}_Y = (R,T,S,P)$. The average payoffs (referred to as scores below) are

$$S_k \equiv \langle \mathbf{S}_k \rangle = \frac{\mathbf{v} \cdot \mathbf{S}_k}{\mathbf{v} \cdot \mathbf{1}} = \frac{D(\mathbf{p},\mathbf{q},\mathbf{S}_k)}{D(\mathbf{p},\mathbf{q},\mathbf{1})} \qquad (4)$$

where $k = X, Y$. Since $\mathbf{v}$ is a stationary, probability vector of the Markov chain, $\mathbf{v} \cdot \mathbf{1} = \mu D(\mathbf{p}, \mathbf{q}, \mathbf{1}) > 0$ for all $\mathbf{p}$ and $\mathbf{q}$. Choosing $\mathbf{p} = \mathbf{q} = \mathbf{1}$ gives $D(\mathbf{1}, \mathbf{1}, \mathbf{1}) = -1$. This means $\mu < 0$ so that $D(\mathbf{p}, \mathbf{q}, \mathbf{1}) < 0$ for all strategy profiles $\{\mathbf{p}; \mathbf{q}\}$. The scores in Eq. 4 can be positive or negative depending on the payoff vectors $\mathbf{S}_k$ so $D(\mathbf{p}, \mathbf{q}, \mathbf{S}_k)$ can be positive or negative. In the case of Axelrod values for the payoff vectors, all scores $S_k$ are positive so $D(\mathbf{p}, \mathbf{q}, \mathbf{S}_k) < 0$.

The transition probability matrix $\mathbf{M} \to \mathbf{M}(\mathbf{p}, \mathbf{q})$ and the accompanying Markov chain correspond to single players X and Y. Each element of a row of $\mathbf{M}$ is linear in one strategy set component, $p_i$ and $q_i$. Assume there is a population of $N$ players each for X and for Y playing the 2X2 game with its own matrix $\mathbf{M}(\mathbf{p}_m, \mathbf{q}_n)$ with different X denoted by index m and Y denoted by index n. Then the average matrix for the population satisfies $\langle \mathbf{M}(\mathbf{p}_m, \mathbf{q}_n) \rangle_{m,n} = \mathbf{M}(\langle \mathbf{p} \rangle, \langle \mathbf{q} \rangle)$ because of the linearity just noted. Below we let $\langle \mathbf{p} \rangle \to \mathbf{p}$ as well as $\langle \mathbf{q} \rangle \to \mathbf{q}$ and make no distinction between only two players X and Y or the averages over populations of players except where noted.

### IV. Strategy Profiles Mapping onto Each Vertex of Stage Game

The vertices of the 2-dimensional representation of the stage game in Fig. 2 are also scores for the IPD. Four sets of strategies, labeled $(xy)$ in the stage game, map onto each of the four vertices, labeled $(S_X, S_Y)$:

$$(cc) \to (R, R); \ (cd) \to (S, T); \ (dc) \to (T, S); \ (dd) \to (P, P).$$

In the IPD, the mapping is much more complex since an infinite number of strategy profiles $\{\mathbf{p}; \mathbf{q}\}$ map onto each vertex. These vertex strategies are summarized as follows:

1. $(R, R)$ Vertex. If $p_1 = q_1 = 1$, then inspection of the first row of the determinant $D(\mathbf{p}, \mathbf{q}, \mathbf{f})$ shows that the first three elements are 0 so $D(\mathbf{p}, \mathbf{q}, \mathbf{f})\big|_{p_1 = q_1 = 1} = -f_1 D_1(\mathbf{p}, \mathbf{q})$. The determinant $D_1(\mathbf{p}, \mathbf{q})$ is the 3X3 determinant obtained from $D(\mathbf{p}, \mathbf{q}, \mathbf{f})$ by removing the first row and fourth column. Then Eq. 4 shows that $S_X = S_Y = R$. This means that an infinite number of strategy profiles $\{\mathbf{p}; \mathbf{q}\}$ with $p_1 = q_1 = 1$ and $D_1(\mathbf{p}, \mathbf{q}) \neq 0$ map onto the $(R, R)$ Vertex.

2. $(P, P)$ Vertex. If $p_4 = q_4 = 0$, then the first three elements of the fourth row are 0 so that $D(\mathbf{p}, \mathbf{q}, \mathbf{f})\big|_{p_4 = q_4 = 0} = f_4 D_4(\mathbf{p}, \mathbf{q})$. The determinant $D_4(\mathbf{p}, \mathbf{q})$ is the 3X3 determinant obtained from $D(\mathbf{p}, \mathbf{q}, \mathbf{f})$ by removing the fourth row and column. Eq. 4 shows that $S_X = S_Y = P$. An infinite number of strategy profiles $\{\mathbf{p}; \mathbf{q}\}$ with $p_4 = q_4 = 0$ and $D_4(\mathbf{p}, \mathbf{q}) \neq 0$ map onto the $(P, P)$ Vertex.

3. $(T,S)$ Vertex. If $p_3 = 0$ and $q_2 = 1$, then the first three elements of the third row are 0 so that $D(\mathbf{p},\mathbf{q},\mathbf{f})\big|_{p_3=0;\; q_2=1} = -f_3 D_3(\mathbf{p},\mathbf{q})$. The determinant $D_3(\mathbf{p},\mathbf{q})$ is the 3X3 determinant obtained from $D(\mathbf{p},\mathbf{q},\mathbf{f})$ by removing the third row and fourth column. Eq. 4 gives $(S_X, S_Y) = (T, S)$, which is infinitely degenerate with respect to the strategy profiles $\{\mathbf{p};\mathbf{q}\}$ with $p_3 = 0$ and $q_2 = 1$ as well as $D_3(\mathbf{p},\mathbf{q}) \neq 0$.

4. $(S,T)$ Vertex. If $p_2 = 1$ and $q_3 = 0$, then the first three elements of the second row are 0 so that $D(\mathbf{p},\mathbf{q},\mathbf{f})\big|_{p_2=1;\; q_3=0} = f_2 D_2(\mathbf{p},\mathbf{q})$. The determinant $D_2(\mathbf{p},\mathbf{q})$ is the 3X3 determinant obtained from $D(\mathbf{p},\mathbf{q},\mathbf{f})$ by removing the second row and fourth column. Eq. 4 give $(S_X, S_Y) = (T, S)$, which is again infinitely degenerate with respect to the strategy set $\{\mathbf{p};\mathbf{q}\}$ with $p_2 = 1$ and $q_3 = 0$ as well as $D_2(\mathbf{p},\mathbf{q}) \neq 0$.

Table 1 summarizes these results for the infinitely degenerate vertex strategy sets.

| Name | Strategy Sets | Stage Game Vertex |
|---|---|---|
| Mutual-Cooperation | $p_1 = q_1 = 1$ with $0 \leq p_i, q_i \leq 1,\ i = 2,3,4$. | $(R,R)$ |
| Mutual-Defection | $p_4 = q_4 = 0$ with $0 \leq p_i, q_i \leq 1,\ i = 1,2,3$. | $(P,P)$ |
| Temptation-Sucker | $p_2 = 0,\ q_3 = 1$ with $0 \leq p_i, q_i \leq 1,\ $ for other $i$ values. | $(T,S)$ |

Table 1. Prisoner's Dilemma Stage Game Vertex Strategy Sets. $(T,S)$ and $(S,T)$ vertices are symmetric with respect to the interchange $\mathbf{p} \leftrightarrow \mathbf{q}$.

### V. Strategies Profiles Mapping onto Edges of the Stage Game

We consider a class of strategy profiles that map onto the four edges of the quadrilateral defined by Fig. 1 and shown as dashed lines in Fig. 2. These are called edge strategy sets. The determinant in Eq. 3 depends linearly on components of the arbitrary vector $\mathbf{f}$ and, hence, on the payoff vectors $\mathbf{S}_X$ and $\mathbf{S}_Y$ when calculating scores using Eq. 4. In addition, two columns depend on the strategy set of just one player, X or Y. These key insights implied that it was possible to select $\{\mathbf{p};\mathbf{q}\}$ so that the 8-dimensional strategy profile mapped onto a line in the 2-dimensional payoff space or, in other words, established a linear correlation between scores. In this section, we introduce the concept of strategy parameter and parameter elimination to show that the strategy sets called ALLC, meaning the player always cooperates, and ALLD, meaning the player always defects, map onto edges of the quadrilateral of the PD stage game. These two strategy sets are of interest in applications of IPD [5,6,9,11,19].

## A. Edge I Strategy Profiles

Along Edge I, the score of X is always greater than that of Y except at the point (R,R). Y selects a fully cooperative strategy set ALLC, $\mathbf{q} \to \mathbf{q}_{ALLC} = (1,1,1,1)$. The determinant in Eq. 3 becomes

$$D(\mathbf{p}, \mathbf{q}_{ALLC}, \mathbf{f}) = \det \begin{vmatrix} -1+p_1 & -1+p_1 & 0 & f_1 \\ p_2 & -1+p_2 & 1 & f_2 \\ p_3 & p_3 & 0 & f_3 \\ p_4 & p_4 & 1 & f_4 \end{vmatrix}. \tag{5}$$

The general expressions for the scores based on Eq. 4 and 5 are $S_X = (R+T\rho)/(1+\rho)$ and $S_Y = (R+S\rho)/(1+\rho)$ where $\rho = (1-p_1)/p_3$ with limits, $0 \leq \rho < \infty$. These expressions for the scores are two parametric equations in terms of the strategy parameter $\rho$ which is dependent on the strategy set chosen by X. Eliminating the strategy parameter gives a linear relation between the scores, $(R-S)S_X + (T-R)S_Y - R(T-S) = 0$, corresponding to Edge I in Eq. 1. The strategy profile $\{\mathbf{p}; \mathbf{q}_{ALLC}\}$ then maps onto Edge I, including the vertices. There are no a priori assumptions regarding the strategy set of X, except that it is a valid conditional transition probability vector. The choice of ALLC by Y is sufficient for scores to fall on Edge I. Here the ratio $\rho = (1-p_1)/p_3$ is considered as a parameter that allows the scores to span their full range as $\rho$ varies over its range 0 to $\infty$. The mapping of the strategy profile $\{\mathbf{p}; \mathbf{q}_{ALLC}\}$ onto Edge I follows immediately.

## B. Edge II Strategy Profiles

Along Edge II, the score for X is again always greater than that for Y except at the point (P,P). If X plays the fully noncooperative strategy set, ALLD, where $\mathbf{p} \to \mathbf{p}_{ALLD} = (0,0,0,0)$, the parametric equations for the scores using Eq. 4 and ALLD for X become $S_X = (T+P\rho)/(1+\rho)$ and $S_Y = (S+P\rho)/(1+\rho)$ with strategy parameter $\rho = (1-q_2)/q_4$ and $0 \leq \rho < \infty$. Eliminating $\rho$ gives the linear relation, $(P-S)S_X + (T-P)S_Y - P(T-S) = 0$ corresponding to Edge II. Along Edge II, $S_X > S_Y$ for all strategy profiles with ALLD for X except those giving (P,P) which occurs when Y defects after mutual defection, $q_4 = 0$. Alternative paths to Edge II occur when X choses the strategy sets $\mathbf{p} = (1,0,0,0)$ and $\mathbf{p} = (0,1,0,0)$ with $0 \leq q_2, q_4 \leq 1$. The scores are the same as Eq. 1 corresponding to Edge II.

## C. Edge IV Strategy Profiles

Along Edge IV, the score of Y is always greater than that of player X except at the point (R,R). The strategic situation for Y along Edge IV is symmetric to that of X along Edge I. Assume X selects a fully cooperative strategy set ALLC, $\mathbf{p} \to \mathbf{p}_{ALLC} = (1,1,1,1)$. Following the analysis for Edge I strategy profiles, the scores are $S_X = (R+S\rho)/(1+\rho)$ and $S_Y = (R+T\rho)/(1+\rho)$ where $\rho = (1-q_1)/q_3$ with limits, $0 \leq \rho < \infty$. Eliminating the strategy parameter $\rho$ gives a linear relation between the scores, $(T-R)S_X + (R-S)S_Y - R(T-S) = 0$, corresponding to Edge IV. The strategy profile $\{\mathbf{p}_{ALLC}; \mathbf{q}\}$ then maps onto Edge IV, including the vertices. There are no a priori assumptions regarding the strategy set for Y, except that it is a valid conditional probability vector.

### D. Edge III Strategy Profiles

Along Edge III, the score for Y is again always greater than that for X except at the point $(P,P)$. The analysis parallels that for Edge II. If Y plays the fully noncooperative strategy, ALLD, where $\mathbf{q} \to \mathbf{q}_{ALLD} = (0,0,0,0)$, then the parametric equations for the scores using Eq. 3 and 4 become $S_X = (S+P\rho)/(1+\rho)$ and $S_Y = (T+P\rho)/(1+\rho)$ with $\rho = (1-p_2)/p_4$ and $0 \leq \rho < \infty$. Eliminating the strategy parameter $\rho$ gives the linear relation, $(T-P)S_X + (P-S)S_Y - P(T-S) = 0$, corresponding to Edge III. $S_Y > S_X$ for all strategy profiles except at $(P,P)$. Alternative paths to Edge III occur if Y choses the strategy sets, $\mathbf{q} = (1,0,0,0)$ and $(0,1,0,0)$. The resulting scores correspond to Edge III.

The strategy sets ALLC and ALLD map onto edges including vertices of the PD stage game. This result is general and independent of the exact strategy set of the other player. Fig. 3 summarizes the conditions for play along the edges of the PD stage game.

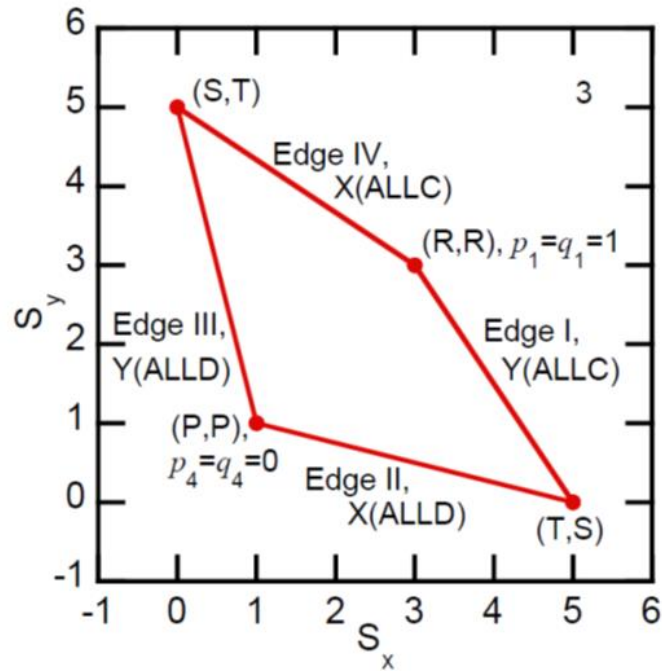

Figure 3. Vertices and Edges of the PD stage game mapped by ALLC and ALLD vs a general strategy set. See text for additional discussion.

10## VI. Mapping onto Interior of Stage Game: TFT, RTFT, WSLS, RWSLS, and $\varepsilon$–Strategy

We leave the vertices and edges of the stage game and venture into the interior of the quadrilateral by selecting more general strategy profiles. Table 2 summarizes the strategy sets that map onto the interior of the PD stage game in the present paper.

| Name (Acronym) | Strategy Sets - Iterated Prisoner's Dilemma | DaMD strategy set | Reference |
|---|---|---|---|
| Tit-for-Tat (TFT) | $(1,0,1,0)$ | Yes | 3, 5,6,8,19 |
| Random Tit-for-Tat (RTFT) | $(1,\nu,1,\nu)$ with $0<\nu<1$ | No | Present paper |
| Generous Tit-for-Tat (GTFT) | $(1,1/3,1,1/3)$ | No | 6 |
| Win-Stay-Lose-Shift (WSLS) | $(1,0,0,1)$ | No | 5,6 |
| Random Win-Stay-Lose-Shift (RWSLS) | $(1,\nu,\nu,1)$ with $0<\nu<1$ | No | Present paper |
| RANDOM | $(1/2,1/2,1/2,1/2)$ | No | 3 |
| $\varepsilon$–Strategy | $(\varepsilon,\varepsilon,\varepsilon,\varepsilon)$ with $0<\varepsilon<1$ | No | Present paper |
| Defect after Mutual Defection (DaMD). Sec. VIII. | $p_4 = 0$ | Yes | Present paper |

Table 2. Stage Game Interior Strategy Sets analyzed in the present paper.

Tit-for-tat (TFT) and win-stay-lose shift (WSLS) are two well-known strategy sets. TFT strategy involves repeating the other player's action on the next step. It is described by the strategy set $\mathbf{p} \to \mathbf{p}_{TFT} = (1,0,1,0)$. A more general strategy set involves letting $\mathbf{p} \to \mathbf{p}_{RTFT} = (1,\nu,1,\nu)$ where the parameter $\nu$ is any rational number satisfying $0<\nu<1$. This is called the random tit-for-tat (RTFT) strategy set. WSLS involves not changing action if both players had the same action on the previous round and changing action if the players had different actions on previous round. It is described by $\mathbf{p} \to \mathbf{p}_{WSLS} = (1,0,0,1)$. A random WSLS strategy set is defined by $\mathbf{p} \to \mathbf{p}_{RWSLS} = (1,\nu,\nu,1)$ and given the acronym, RWSLS.

A third random strategy set called the $\varepsilon$–strategy is defined by $\mathbf{q} \to \mathbf{q}_\varepsilon = \varepsilon(1,1,1,1)$ where $\varepsilon$ is a rational number satisfying $0 \le \varepsilon \le 1$. This strategy is Axelrod's RANDOM strategy when $\varepsilon = 1/2$ [3]. The $\varepsilon$–strategy provides a mechanism for nonlinear trajectories of scores in the interior of the stage game. The parameter $\varepsilon$ in the $\varepsilon$–strategy means that the player cooperates with a probability $\varepsilon$ and defects with $1-\varepsilon$. The value $\varepsilon = 1/2$ is considered a "coin-flipping" strategy set for a player who posits that unpredictability can result in an enhanced score or who has no rational model of the IPD but must compete. Since $\varepsilon = \langle q_i \rangle$ taken over an ensemble, there can be fluctuations in the individual $q_i$ in a stationary Markov chain. For example, the $\varepsilon = 0.95$ can correspond to the average $\langle q_i \rangle$ for a player intending to play ALLC but making errors.

### A. TFT and RTFT vs $\varepsilon$–Strategy

With the strategy profile, $\mathbf{p} \to \mathbf{p}_{\text{RTFT}} = (1, v, 1, v)$ and $\mathbf{q} \to \mathbf{q}_\varepsilon = \varepsilon(1,1,1,1)$, the determinant in Eq. 3 becomes

$$D(\mathbf{p}_{\text{RTFT}}, \mathbf{q}_\varepsilon, \mathbf{f}) = -\varepsilon^2 f_1 - \varepsilon(1-\varepsilon)\left[v f_1 + f_2 + (1-v) f_3\right] - (1-\varepsilon)^2 \left[v f_2 + (1-v) f_4\right]. \quad (6)$$

The scores are $S_X(v, \varepsilon) = \varepsilon^2 R + (1-\varepsilon)^2 \left[v S + (1-v) P\right] + \varepsilon(1-\varepsilon)\left[S + v R + (1-v) T\right]$ and $S_Y(v, \varepsilon) = \varepsilon^2 R + (1-\varepsilon)^2 \left[v T + (1-v) P\right] + \varepsilon(1-\varepsilon)\left[T + v R + (1-v) S\right]$. These scores are parametric equations in the two strategy parameters $v$ and $\varepsilon$ which determine the strategy profile $\{\mathbf{p}_{\text{RTFT}}; \mathbf{q}_\varepsilon\}$. The scores are linear in $v$ but nonlinear in $\varepsilon$ because each $q_i = \varepsilon$. We consider two cases obtained by holding one parameter constant and varying the other.

#### 1. Constant $v$ (RTFT) and varying $\varepsilon$ ($\varepsilon$–Strategy)

TFT has $v = 0$, $\mathbf{p}_{\text{TFT}} = (1, 0, 1, 0)$, and $S_X(0, \varepsilon) = S_Y(0, \varepsilon) = \varepsilon^2 R + (1-\varepsilon)^2 P + \varepsilon(1-\varepsilon)(T+S)$. The scores are equal and trace out the line from $(P, P)$ to $(R, R)$ as $\varepsilon$ varies between 0 and 1. The case $v = 1$ gives ALLC with $\mathbf{p}_{\text{ALLC}} = (1,1,1,1)$. Scores are $S_X(1, \varepsilon) = \varepsilon R + (1-\varepsilon) S$, and $S_Y(1, \varepsilon) = \varepsilon R + (1-\varepsilon) T$. The strategy parameter $\varepsilon$ can be eliminated to give $(T-R) S_X + (R-S) S_Y - R(T-S) = 0$, corresponding to Edge IV, including the mutual Pareto optimum $(R, R)$ of the stage game as well as $(S, T)$. The Generous Tit-for Tat (GTFT) has $v = 1/3$ with scores $S_X(1/3, \varepsilon) = \varepsilon^2 R + (1-\varepsilon)^2 (2P + S)/3 + \varepsilon(1-\varepsilon)(2T + R + 3S)/3$ and $S_X(1/3, \varepsilon) = \varepsilon^2 R + (1-\varepsilon)^2 (2P + T)/3 + \varepsilon(1-\varepsilon)(3T + R + 2S)/3$. These parametric equations in $\varepsilon$ for GTFT correspond to a curve with endpoints $(R, R)$ for $\varepsilon = 1$ and $\left[(2P+S)/3, (2P+T)/3\right]$ for $\varepsilon = 0$. The endpoint for $\varepsilon = 0$ falls on Edge III. Finally, the RANDOM strategy with $v = 1/2$ gives the following scores, $S_X(1/2, \varepsilon) = \varepsilon^2 R + (1-\varepsilon)^2 (P+S)/2 + \varepsilon(1-\varepsilon)(T + R + 2S)/2$, and $S_Y(1/2, \varepsilon) = \varepsilon^2 R + (1-\varepsilon)^2 (P+T)/2 + \varepsilon(1-\varepsilon)(2T + R + S)/2$. These parametric equations in $\varepsilon$ correspond to a curve with endpoints $(R, R)$ for $\varepsilon = 1$ and $\left[(P+S)/2, (T+P)/2\right]$ for $\varepsilon = 0$. The endpoint for $\varepsilon = 0$ again falls on Edge III. Fig. 4 summarizes these results.

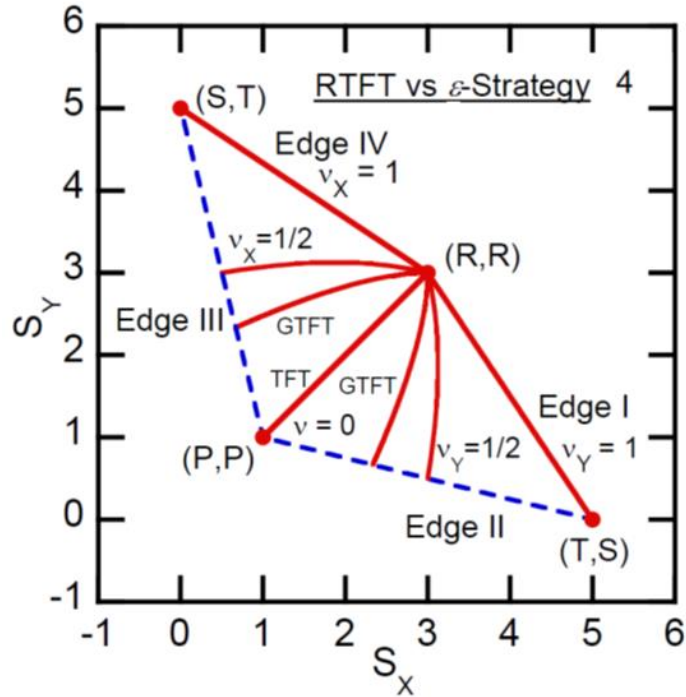

Figure 4. RTFT vs $\varepsilon$–Strategy. X playing TFT ($\nu = 0$) gives the line from (P,P) to (R,R). X playing GTFT ($\nu_X = 1/3$) shows curvature and is in region $S_X < S_Y$. See text.

The selection of the TFT strategy set ($\nu = 0$) by X results in equal scores for X and Y. But if X selects any other value of $\nu > 0$, the result is a score that falls in the region where $S_Y > S_X$, including the GTFT strategy set. We return to the issue of score orderings below. Although cases of TFT and ALLC result in straight lines, the other cases of $\nu$, for example GTFT, are quadratic in $\varepsilon$ and are curvilinear in Fig. 4. Since the IPD is symmetric in $X \leftrightarrow Y$, similar results follow for the region where $S_X > S_Y$. These lines are also shown in Fig. 4. The dashed edges are not traced by a single strategy profile but are the endpoints for (infinitely many) lines and curves traced by the strategy profiles with constant $\nu$.

### 2. Constant $\varepsilon$ ($\varepsilon$–Strategy) and varying $\nu$ (RTFT)

The case $\varepsilon = 1$ gives ALLC for Y so $\mathbf{q} \to \mathbf{q}_{ALLC} = (1,1,1,1)$. Section V.A predicts this strategy set for Y maps into Edge I, and for $\varepsilon = 1$ the scores for are indeed equal, $S_X = S_Y = R$, the endpoint for Edge I. The case $\varepsilon = 0$ gives ALLD for Y so $\mathbf{q} \to \mathbf{q}_{ALLD} = (0,0,0,0)$ and Sec. V.D predicts that this strategy set maps to Edge III. The scores are $S_X(\nu, \varepsilon = 0) = \nu S + (1-\nu)P$, and $S_Y(\nu, \varepsilon = 0) = \nu T + (1-\nu)P$. The parameter $\nu$ can be eliminated to give $(T-P)S_X + (P-S)S_Y - P(T-S) = 0$, the general relation for Edge III including $(P,P)$ and $(T,S)$ of the PD stage game. The case $\varepsilon = 1/2$ gives the strategy profile RTFT vs. RANDOM with scores, $S_X(\nu, \varepsilon = 1/2) = \left[(T+R+P+S) - \nu(T+P-R-S)\right]/4$, and $S_Y(v,\varepsilon=1/2) = [(T+R+P+S)+v(T+R-P-S)]/4$. The strategy parameter $v$ can be eliminated to give $2(T+R-P-S)S_X + 2(T+P-R-S)S_Y - (T-S)(T+R+P+S) = 0$, a line with endpoints $S_X = S_Y = (T+R+P+S)/4$ and $[(R+S)/2, (T+R)/2]$, which lies on Edge IV. For Axelrod values, $14S_X + 6S_Y - 45 = 0$. Fig. 5 summarizes these results for RTFT vs $\varepsilon$–Strategy.

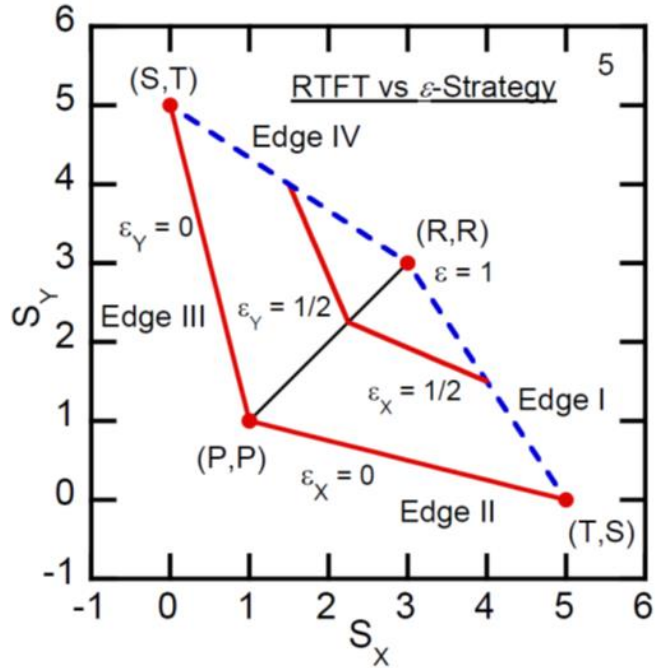

Figure 5. RTFT vs $\varepsilon$–Strategy. X playing ALLD ($\varepsilon_X = 0$) gives Edge II. Y playing ALLD gives Edge III. X or Y choosing $\varepsilon = 0$ gives $(R,R)$. See text.

The lines and curves in Fig. 4 and 5 cover the vertices, edges, and interior of the Prisoner's Dilemma stage game as the parameters range through the rational numbers from 0 to 1.

### B. TFT vs General Y Strategy

With the general strategy profile $\{\mathbf{p}_{TFT}; \mathbf{q}\}$, the determinant in Eq. 3 becomes $D(\mathbf{p}_{TFT}, \mathbf{q}, \mathbf{f}) = -f_1 q_2 q_4 - (f_2 + f_3)(1-q_1)q_4 - f_4(1-q_1)(1-q_3)$. Using Eq. 4, the scores are $S_X = S_Y = [R + (T+S)\rho_1 + P\rho_1\rho_3]/[1 + 2\rho_1 + \rho_1\rho_3]$ where the two strategy parameters are given by $\rho_1 = (1-q_1)/q_2$, and $\rho_3 = (1-q_3)/q_4$ with the conditions $0 \leq \rho_i \leq \infty$. The scores map onto the line from the endpoint $(P,P)$ for $\rho_1$ finite and $\rho_3 \to \infty$ $(q_4 = 0)$ to the endpoint $(R,R)$ for $\rho_1 = \rho_3 = 0$ $(q_1 = q_3 = 1)$, as expected. The line in Fig. 4 from $(P,P)$ to $(R,R)$, therefore, also describes TFT vs General Y Strategy.

## C. WSLS and RWSLS vs $\varepsilon$–Strategy

With these strategy profiles, $\{\mathbf{p}_{\text{RWSLS}}; \mathbf{q}_\varepsilon\}$, the determinant in Eq. 3 becomes

$$D(\mathbf{p}_{\text{RWSLS}}, \mathbf{q}_\varepsilon, \mathbf{f}) = -v\varepsilon^2 f_1 - (1-\varepsilon)^2 \left[ f_2 + (1-v)f_4 \right] - \varepsilon(1-\varepsilon)\left[ f_1 + vf_2 + (1-v)f_3 \right]. \quad (7)$$

The scores are obtained from Eq. 7 as follows:

$$D(\mathbf{p}_{\text{RWSLS}}, \mathbf{q}_\varepsilon, \mathbf{S}_X) = -v\varepsilon^2 R - (1-\varepsilon)^2 \left[ S + (1-v)P \right] - \varepsilon(1-\varepsilon)\left[ R + vS + (1-v)T \right], \quad (8)$$

$$D(\mathbf{p}_{\text{RWSLS}}, \mathbf{q}_\varepsilon, \mathbf{S}_Y) = -v\varepsilon^2 R - (1-\varepsilon)^2 \left[ T + (1-v)P \right] - \varepsilon(1-\varepsilon)\left[ R + vT + (1-v)S \right], \quad (9)$$

$$D(\mathbf{p}_{\text{RWSLS}}, \mathbf{q}_\varepsilon, \mathbf{1}) = -2(1-\varepsilon) - v\left[ \varepsilon^2 - (1-\varepsilon)^2 \right]. \quad (10)$$

Then, the scores from Eq. 4 are

$$S_X(v,\varepsilon) = \frac{D(\mathbf{p}_{\text{RWSLS}}, \mathbf{q}_\varepsilon, \mathbf{S}_X)}{D(\mathbf{p}_{\text{RWSLS}}, \mathbf{q}_\varepsilon, \mathbf{1})}, \quad S_Y(v,\varepsilon) = \frac{D(\mathbf{p}_{\text{RWSLS}}, \mathbf{q}_\varepsilon, \mathbf{S}_Y)}{D(\mathbf{p}_{\text{RWSLS}}, \mathbf{q}_\varepsilon, \mathbf{1})}. \quad (11)$$

The case $v = 0$ corresponds the WSLS playing against the $\varepsilon$–strategy. These scores are again parametric equations in the two strategy parameters $v$ and $\varepsilon$ which determine the strategy profile $\{\mathbf{p}_{\text{RTFT}}; \mathbf{q}_\varepsilon\}$. The scores are linear in $v$ but nonlinear in $\varepsilon$ because each $q_i = \varepsilon$.

The case $v = 0$ gives the classic WSLS strategy set $\mathbf{p}_{\text{WSLS}} = (1,0,0,1)$ vs the $\varepsilon$–strategy. Eq. 11 gives $S_X(0,\varepsilon) = \left[\varepsilon(R+T) + (1-\varepsilon)(P+S)\right]/2$, and $S_Y(0,\varepsilon) = \left[\varepsilon(R+S) + (1-\varepsilon)(T+P)\right]/2$. The strategy parameter $\varepsilon$ is eliminated to give the linear relation for $\{\mathbf{p}_{\text{WSLS}}; \mathbf{q}_\varepsilon\}$, $2(T+P-R-S)S_X + 2(T+R-P-S)S_Y - (T+R)(T+P) + (R+S)(P+S) = 0$, with endpoints $\left[(R+T)/2, (R+S)/2\right]$ and $\left[(P+S)/2, (T+P)/2\right]$ lying on Edges I and III, respectively. For Axelrod values, the line becomes $6S_X + 14S_Y - 45 = 0$ with endpoints $(1/2, 3)$ and $(4, 3/2)$. It is labeled $v_X = 0$, WSLS, in Fig. 6. This line shows that the strategy profile $\{\mathbf{p}_{\text{WSLS}}; \mathbf{q}_\varepsilon\}$ results in scores that that are larger for X, $S_X > S_Y$, when $\varepsilon > 1/2$, and for Y, $S_Y > S_X$, when $\varepsilon < 1/2$. The case of equal scores $S_X = S_Y = (T+R+P+S)/4$ occurs when $\varepsilon = 1/2$. This ordering of scores is discussed further in the next section with Y playing a general strategy set. The line for Y playing WSLS and X playing the $\varepsilon$–strategy is also shown. This line has endpoints on Edges II and IV and passes through the point of equal scores as shown in Fig. 6.

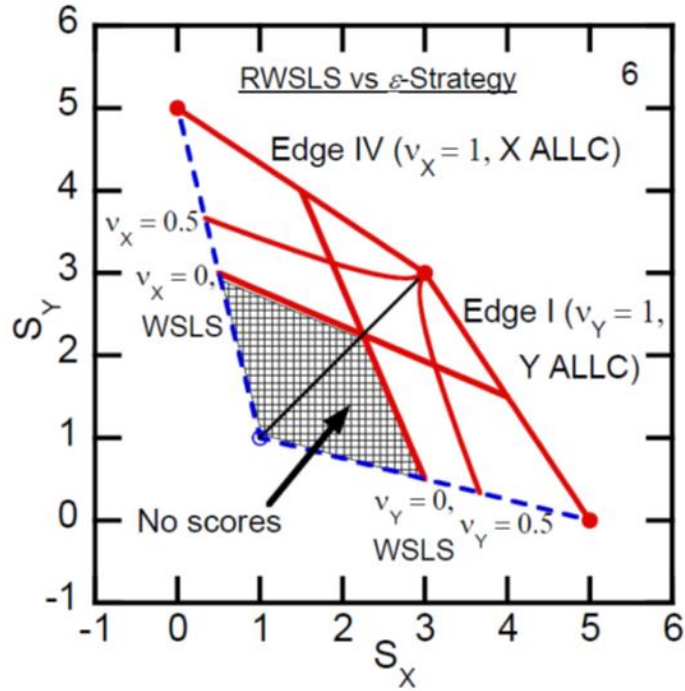

*Figure 6. RWSLS vs ε-Strategy. X playing ALLC gives Edge IV. X playing WSLS (vx = 0) gives line from Edge III to Edge I. Symmetric lines for Y also shown. No scores are allowed in cross hatched region including (P,P).*

The case $v_X = 1$, ALLC for X, maps into Edge IV as expected. The case for $v_X = 1/2$ gives $\mathbf{p}_{\text{RWSLS}} = (1, 1/2, 1/2, 1)$ so that scores are given by Eq. 8 to 11 as

$$S_X(1,\varepsilon) = \frac{\varepsilon^2 R/2 + \varepsilon(1-\varepsilon)\left[R + (T+S)/2\right] + (1-\varepsilon)^2 (S + P/2)}{2(1-\varepsilon) + \left[\varepsilon^2 - (1-\varepsilon)^2\right]/2}, \tag{12}$$

$$S_Y(1,\varepsilon) = \frac{\varepsilon^2 R/2 + \varepsilon(1-\varepsilon)\left[R + (T+S)/2\right] + (1-\varepsilon)^2 (T + P/2)}{2(1-\varepsilon) + \left[\varepsilon^2 - (1-\varepsilon)^2\right]/2}. \tag{13}$$

The parametric Eq. 12 and 13 are quadratic in the strategy parameter $\varepsilon$ so curvature is expected and clearly observed in Fig. 6 for the curvilinear line labeled $v_X = 1/2$. The endpoints of the curvilinear line are $(R, R)$ for $\varepsilon = 1$ and, for $\varepsilon = 0$, $\left[(2S+P)/3, (2T+P)/3\right]$ which lies on Edge III. All scores for $v_X = 1/2$ satisfy $S_Y > S_X$ except for $(R,R)$. There is $X \leftrightarrow Y$ symmetry for playing RWSLS vs $\varepsilon$–strategy so comparable lines are plotted in Fig. 6. There are no scores allowed in the quadrilateral bounded by the two lines for WSLS vs $\varepsilon$–strategy and Edges II, III. For $\varepsilon = 0$, scores fall on Edges II and III, but are limited by endpoints of WSLS lines from approaching $(P, P)$.

## D. WSLS vs General Strategy

With the strategy profile $\{\mathbf{p}_{\text{WSLS}}; \mathbf{q}\}$, the determinant in Eq. 3 becomes $D(\mathbf{p}_{\text{WSLS}}, \mathbf{q}, \mathbf{f}) = -f_1(1-q_2)q_4 - (f_2+f_4)(1-q_1)(1-q_2) - f_3 q_3(1-q_1)$. Eq. 4 gives the scores

$$S_X = \frac{T\rho_1 + R\rho_2 + (S+P)\rho_1\rho_2}{\rho_1 + \rho_2 + 2\rho_1\rho_2}, \quad S_Y = \frac{S\rho_1 + R\rho_2 + (T+P)\rho_1\rho_2}{\rho_1 + \rho_2 + 2\rho_1\rho_2}. \tag{14}$$

The two strategy parameters are $\rho_1 = (1-q_1)/q_4$ and $\rho_2 = (1-q_2)/q_3$ with the conditions $0 \le \rho_i \le \infty$. The parametric equations give essentially an infinite number of linear relations between the scores for various conditions on the two strategy parameters, and hence, the transition probability vector $\mathbf{q}$. If $\rho_1 = 0$ ($q_1 = 1$), then $S_X = S_Y = R$. If $\rho_2 = 0$ ($q_2 = 1$), then $S_X = T$, $S_Y = S$. If $\rho_1 \to \infty$ ($q_4 = 0$), then the scores are $S_X = [T + \rho_2(P+S)]/(1+2\rho_2)$ and $S_Y = [S + \rho_2(T+P)]/(1+2\rho_2)$. The endpoints of the line are $[(P+S)/2, (T+P)/2]$ on Edge III and $(T,S)$. The strategy parameter $\rho_2$ can be eliminated between the last two equations for the scores to get: $(T+P-2S)S_X + (2T-P-S)S_Y - T(T+P) + S(S+P) = 0$. Axelrod values give $2S_X + 3S_Y - 10 = 0$. As a final example, let $\rho_2 = 1$ ($q_2 + q_3 = 1$) so that $S_X = S_Y = [R + \rho_1(T+P+S)]/(1+3\rho_1)$. These equal scores form a line with endpoints $(R,R)$ when $\rho_1 = 0$ ($q_1 = 1$) and $S_X = S_Y = (T+P+S)/3$ when $\rho_1 \to \infty$ ($q_4 = 0$). Plots of these general results give a figure like Fig. 6 with the addition of the line for equal scores from $(R,R)$ to $S_X = S_Y = (T+P+S)/3$. For Axelrod values, these endpoints are $(3,3)$ to $(2,2)$. Varying the two strategy parameters generates an infinity of lines covering the area as indicated in Fig. 6, but lines are excluded from the cross hatched area with vertex now at $(2,2)$.

Fig. 6 shows that WSLS vs $\varepsilon$–strategy falls along a line that satisfies three orderings $S_X > S_Y$, $S_Y > S_X$, and $S_X = S_Y$. For X choosing WSLS and Y choosing a general strategy, the relative magnitudes of the scores satisfy the following similar conditions:

i) $S_X > S_Y$ (altruistic): This gives $T\rho_1 + R\rho_2 + (S+P)\rho_1\rho_2 > S\rho_1 + R\rho_2 + (T+P)\rho_1\rho_2$. If $\rho_1 > 0$ ($q_1 \ne 1$ and $q_4 \ne 0$) so it can be eliminated in the condition, then $\rho_2 < 1$, and the inequality in the scores results from $q_2 + q_3 > 1$.

ii) $S_Y > S_X$ (extortive): This gives $T\rho_1 + R\rho_2 + (S+P)\rho_1\rho_2 < S\rho_1 + R\rho_2 + (T+P)\rho_1\rho_2$. Similar analysis shows that the inequality in the scores results from $q_2 + q_3 < 1$.

iii) $S_X = S_Y$ (reciprocal): The equality in scores clearly results from $q_2 + q_3 = 1$.

If X chooses WSLS, then the Y strategy set determines the orderings (extortive, reciprocal, and altruistic) of the scores. Stated another way, Y choosing a strategy set that can be extortive, reciprocal, or altruistic is conditional on X choosing WSLS. Note here that there is no condition on $q_4$, the probability to defect after mutual defection. These results are consistent with those when X choses WSLS and Y choses the $\varepsilon$–strategy discussed in the previous section. More comments on this surprising result are in Section IX. Conclusions.

## VII. DaMD Strategy Set vs ε-Strategy and WSLS. Reciprocity and Extortion

The DaMD strategy set is defined by $\mathbf{p}_{\text{DaMD}} = (p_1, p_2, p_3, p_4 = 0)$. For the strategy profile $\{\mathbf{p}_{\text{DaMD}}, \mathbf{q}_\varepsilon\}$, Eq. 3 gives

$$D(\mathbf{p}_{\text{DaMD}}, \mathbf{q}_\varepsilon, \mathbf{f}) = -p_3\{[\varepsilon^2 f_1 + \varepsilon(1-\varepsilon)f_2] + [\varepsilon\rho_1 + (1-\varepsilon)\rho_2][\varepsilon f_3 + (1-\varepsilon)f_4]\}, \quad (14)$$

with two strategy parameters are defined as $\rho_1 = (1-p_1)/p_3$, and $\rho_2 = (1-p_2)/p_3$. Eq. 4 and 14 gives the following expressions for the scores, $S_X = (a+b\rho)/(\varepsilon+\rho)$ and $S_Y = (c+d\rho)/(\varepsilon+\rho)$. The strategy parameter $\rho$ is expressed as $\rho \equiv \varepsilon\rho_1 + (1-\varepsilon)\rho_2$ so $\varepsilon + \rho > 0$. The symbols, $a$, $b$, $c$, $d$, are given by

$$\left.\begin{aligned} a &= \varepsilon^2 R + \varepsilon(1-\varepsilon)S \\ b &= \varepsilon T + (1-\varepsilon)P \\ c &= \varepsilon^2 R + \varepsilon(1-\varepsilon)T \\ d &= \varepsilon S + (1-\varepsilon)P \end{aligned}\right\}. \quad (15)$$

The strategy parameter $\rho$ can be eliminated from the scores giving the linear relation,

$$\alpha(\varepsilon)S_X + \beta(\varepsilon)S_Y + \gamma(\varepsilon) = 0, \quad (16)$$

where

$$\left.\begin{aligned} \alpha(\varepsilon) &= \varepsilon(R-S) + (1-\varepsilon)(T-P) \\ \beta(\varepsilon) &= \varepsilon(T-R) + (1-\varepsilon)(P-S) \\ \gamma(\varepsilon) &= -(T-S)\left[\varepsilon^2 R + \varepsilon(1-\varepsilon)(T+S) + (1-\varepsilon)^2 P\right] \end{aligned}\right\}. \quad (17)$$

When Y plays ALLD so that $\varepsilon = 0$, then the scores map onto $(P, P)$. If Y plays ALLC so that $\varepsilon = 1$, then the linear relationship between scores is $(R-S)S_X + (T-R)S_Y - R(T-S) = 0$, giving Edge I. If Y plays RANDOM, $\varepsilon = 1/2$, then the scores satisfy a linear relationship in the interior of the stage game, $2(T+R-P-S)S_X + 2(T+P-R-S)S_Y - (T-S)(T+R+P+S) = 0$. For Axelrod values, the line is $14S_X + 6S_Y - 45 = 0$. The end points of this line fall on Edges II and IV. There is symmetry with player Y playing DaMD and X playing an $\varepsilon$–strategy. For this strategy profile, the line for $\varepsilon = 1$ for X maps onto Edge IV; $\varepsilon = 0$ maps onto $(P, P)$. The line with $\varepsilon = 1/2$ maps onto the interior of the stage game quadrilateral. These lines are shown in Fig. 7.

As $\varepsilon$ scans between its limits, the edges and interior of the quadrilateral are covered by the lines generated.

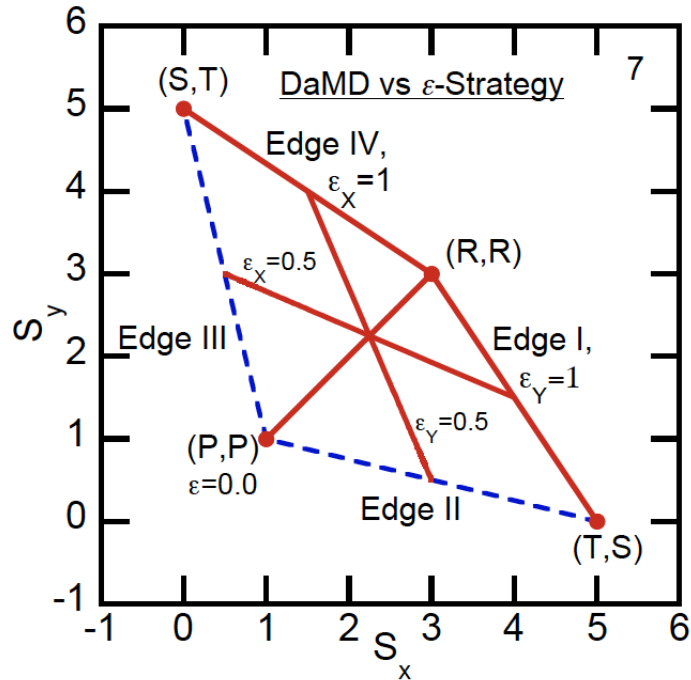

Figure 7. DaMD vs $\varepsilon$-Strategy. X plays DaMD: Y plays ALLC ($\varepsilon = 1$) giving Edge I; Y plays ALLD ($\varepsilon = 0$) giving (P,P). Line from (P,P) to (R,R) described in Sec. VIII. Symmetric results for X $\leftrightarrow$ Y.

Fig. 7 shows that the line generated when X plays DaMD and Y plays RANDOM falls in the regions where $S_X > S_Y$ and $S_X < S_Y$. The exact point on the line depends on the value of the strategy parameter $\rho$ through $\rho_1$ and $\rho_2$, so ultimately through the probabilities $p_i$, $i = 1,2,3$. There are two orderings of the scores other than equal scores:

i) $S_X > S_Y$ (extortive): This condition gives $\rho > 1-\varepsilon$ so that $\varepsilon(1-p_1) > (1-\varepsilon)(p_2 + p_3 - 1)$. The inequality is valid for arbitrary $\varepsilon$ and $p_1$, with $\varepsilon$, $1-\varepsilon$, $1-p_1 \neq 0$, and $p_2 + p_3 < 1$.

ii) $S_Y > S_X$ (altruistic): This condition gives $\rho < 1-\varepsilon$ which in turn gives $\varepsilon(1-p_1) < (1-\varepsilon)(p_2 + p_3 - 1)$. If $\varepsilon$, $1-\varepsilon$, $1-p_1 \neq 0$, then $p_2 + p_3 > 1$.

If Y chooses to play WSLS rather than a $\varepsilon$–strategy, Eq. 3 gives for the strategy profile $\{\mathbf{p}_{\text{DaMD}}, \mathbf{q}_{\text{WSLS}}\}$ and $D(\mathbf{p}_{\text{DaMD}}, \mathbf{q}_{\text{WSLS}}, \mathbf{f}) = -(1 - p_1)[p_3 f_2 + (1 - p_2)(f_3 + f_4)]$. Then Eq. 4 gives the scores as $S_X = [S+(T+P)\rho]/(1+2\rho)$ and $S_Y = [T+(S+P)\rho]/(1+2\rho)$ with the strategy parameter given by $\rho = (1-p_2)/p_3$ where $0 \leq \rho \leq \infty$. The strategy parameter can be eliminated to get the line, $(2T-P-S)S_X + (T+P-2S)S_Y - T(T+P) + S(S+P) = 0$. For Axelrod values, $3S_X + 2S_Y - 10 = 0$ for DaMD vs WSLS. The endpoints of the line are the vertex $(S,T)$ and $[(T+P)/2, (S+P)/2]$ on Edge II. There are again two possibilities for the order of the scores:

i)         $S_X > S_Y$ (extortive): This gives $\rho > 1$ and $p_2 + p_3 < 1$.
ii)       $S_Y > S_X$ (altruistic): This gives $\rho < 1$ and $p_2 + p_3 > 1$.

The terms extortive and altruistic for Y playing WSLS rather than X are used in an opposite fashion to the previous section because here the strategy of player X determines the orderings of scores. The condition $p_2 + p_3 < 1$ for ensuring $S_X > S_Y$, is a general property of DaMD not limited to Y playing an $\varepsilon$–strategy or WSLS. We show this in detail in the next Section.

## VIII. DaMD vs Arbitrary Strategy

We let X play DaMD and Y play an arbitrary strategy. Eq. 3 gives

$$D(\mathbf{p}_{\text{DaMD}}, \mathbf{q}, \mathbf{f}) = -\sum_{i=1}^{4} f_i D_i(\mathbf{p}_{\text{DaMD}}, \mathbf{q}), \tag{18}$$

where $D_{1,3}(\mathbf{p}_{\text{DaMD}}, \mathbf{q}) = q_4 \delta_{1,3}(\mathbf{p}_{\text{DaMD}}, \mathbf{q})$, $D_2(\mathbf{p}_{\text{DaMD}}, \mathbf{q}) = -q_4 \delta_2(\mathbf{p}_{\text{DaMD}}, \mathbf{q})$, and $D_4 = (1-q_1)\delta_1 + q_3 \delta_2 + (1-q_2)\delta_3$. The 2X2 determinants $\delta_i(\mathbf{p}_{\text{DaMD}}, \mathbf{q})$ are

$$\delta_1 = \det\begin{vmatrix} p_2 q_3 & -1+p_2 \\ p_3 q_2 & p_3 \end{vmatrix}, \quad \delta_2 = \det\begin{vmatrix} -1+p_1 q_1 & -1+p_1 \\ p_3 q_2 & p_3 \end{vmatrix}, \quad \delta_3 = \det\begin{vmatrix} -1+p_1 q_1 & -1+p_1 \\ p_2 q_3 & -1+p_2 \end{vmatrix}. \tag{19}$$

Eq. 4, 18, and 19 give the scores, $S_X$ and $S_Y$. If $S_X > S_Y$, then it is readily seen that $D(\mathbf{p}_{\text{DaMD}}, \mathbf{q}, \mathbf{S}_X) < D(\mathbf{p}_{\text{DaMD}}, \mathbf{q}, \mathbf{S}_Y)$ since $D(\mathbf{p}_{\text{DaMD}}, \mathbf{q}, \mathbf{1}) < 0$ in general. For the symmetric PD stage game, $f_1 = R$ and $f_4 = P$. This means that $S\delta_2 - T\delta_3 < T\delta_2 - S\delta_3$ after $q_4 \neq 0$ has been eliminated. This means $-\delta_2 < \delta_3$ if $q_4 \neq 0$. Evaluating the determinants, substituting, and rearranging results in the inequality $p_2 + p_3 - 1 < (1-p_1)(p_2 q_3 + p_3 q_2)/(1-p_1 q_1)$. This inequality is true for all $q_i$ if $p_2 + p_3 < 1$ and ensures that $S_X > S_Y$ so this is an extortive DaMD since player X determines the score orderings.

Fig. 7 also shows the line from $(P,P)$ to $(R,R)$ corresponding to the case of equal scores, $S_Y = S_X$. In the case of equal scores for DaMD vs $\varepsilon$-strategy, we have $a+b\rho = c+d\rho$ from Sec. VII. This condition means that $\rho = 1 - \varepsilon$. From the definition of the strategy parameter $\rho$, the following condition on the $p_i$ follows, $\varepsilon(1-p_1) = (1-\varepsilon)(p_2 + p_3 - 1)$. This condition is true for all $\varepsilon$ if $p_2 + p_3 = 1$, $p_1 = 1$. The scores now read $S_X = S_Y = \varepsilon^2 R + \varepsilon(1-\varepsilon)(T+S) + (1-\varepsilon)^2 P$ and give the line from $(P,P)$ to $(R,R)$ as $\varepsilon$ ranges from 0 to 1. This result can also be generalized. The condition $p_2 + p_3 = 1$ for ensuring $S_X = S_Y$ is a general property of DaMD not limited to Y playing an $\varepsilon$–strategy. To see this, we use the same argument as above to show that $S_X = S_Y$ leads to the condition $-\delta_2 = \delta_3$, implying that $-(1-p_1q_1)(1-p_2-p_3) = (1-p_1)(p_2q_3 + p_3q_2)$. This condition is valid for all $\mathbf{q}$ when $p_1 = 1$, and $p_2 + p_3 = 1$. With these conditions on the $p_i$, Eq. 4 and 18 give the line from $(P,P)$ to $(R,R)$ as $S_X = S_Y = (R + P\rho)/(1+\rho)$ where the strategy parameter is now $\rho = [(1-q_1)/q_4]\{[(1-q_2) + p_3(q_2 + q_3)]/[(1-p_3)q_3 + p_3q_2]\}$ and $0 \leq \rho \leq \infty$ for all $\mathbf{q}$ and $p_3$. $S_X = S_Y = R$ when $q_1 = 1$ and $\rho = 0$. Also $S_X = S_Y = P$ when $q_4 = 0$ and $\rho \to \infty$. Again, the strategy parameter analysis allows a more transparent analysis of the linear relation between scores. The strategy parameter is a function of the strategy profile $\rho = \rho(\mathbf{p}, \mathbf{q})$ so as the strategy profile $\{\mathbf{p};\mathbf{q}\}$ varies, the strategy parameter $\rho$ scans through its limits $0 \leq \rho \leq \infty$ and traces through the linear relation for the scores. Table 3 summarizes the results for DaMD vs an arbitrary strategy set.

| DaMD Strategy for player X | Score Ordering | Conditions on probabilities |
| --- | --- | --- |
| Generic strategy | Not specified | $p_4 = 0$ |
| Extortive strategy | $S_X > S_Y$ | $p_4 = 0$, $p_2 + p_3 < 1$ |
| Reciprocal strategy | $S_X = S_Y$ | $p_1 = 1$, $p_4 = 0$, $p_2 + p_3 = 1$ |

Table 3. DaMD strategies with generic, extortive, and reciprocal conditions.

## IX. Conclusions

The general analysis presented here establishes many specific features of the mapping from the 8-dimensioal hypercube defined by the strategy profile $\{\mathbf{p};\mathbf{q}\}$ onto the 2-dimensional space of scores $(S_X, S_Y)$. The concept of strategy parameter is key to this technique to identify the lines that are mapped by a strategy profile. Vertex strategy profiles are shown in Fig. 2 and Table 1. Edge strategy profiles are described in Fig. 3 and Sec. V. These ALLC and ALLD are the key edge strategy sets with ALLC mapping onto Edges I and IV and ALLD mapping onto Edges II and III for any opponent strategy sets. In the specific example of Ref. 8, the extortionate strategy set (ZDS) is played against an ALLC strategy so maps onto Edge I. This is readily seen by taking Eq. 16 in Ref. 8 and eliminating the parameter $\chi$ resulting in the expression for Edge I. Press and Dyson also showed how a player Y, whom they call "evolutionary" and who only acts to increase Y's own score without regard for the score of X, can be extorted if X plays the ZDS. We refer to the "evolutionary" player who only considers that player's own score as a "naïve" player. We return to this feature for DaMD below.

Several famous strategy profiles are shown to map onto lines in the interior of the PD stage game. The $\varepsilon$-strategy is played against tit-for-tat (TFT) and win-stay-lose-shift (WSLS) as well as random versions of these two strategies (RTFT and RWSLS) as models. TFT and WSLS are also played against a general strategy set for Y. The scores for these games are expressed in terms of one or more strategy parameters which depend on a specific combination of the transition probabilities, $\mathbf{p}$ or $\mathbf{q}$, appropriate to the strategy profiles. The strategy parameter(s) is then eliminated between the scores for both players resulting in a linear relation between the scores. This is done for mapping of the 8-dimensional hypercube of strategy profiles into the 2-dimensional space of scores including edges and the interior of the stage game. The specific expressions for scores give the endpoints of the line. Linear relations between scores are ubiquitous and are not specific to ZDS or DaMD. The scores for one player acting on the $\varepsilon$–strategy are quadratic in $\varepsilon$ so the trajectories of the scores obtained by varying the strategy parameter $\varepsilon$ show curvature as in Figs. 4 and 6. The lines and curves give insight into the behavior of the scores as the strategy profiles are systematically varied. Recently human-machine experiments have verified predictions of the linear relations between scores [18,20]. Our analysis gives many new predictions that can guide experiment and decision making.

It is important to be clear on the concept of a strategy set that can extort the other player. The first specific example with an extortive score condition in the present work came when X chose WSLS while Y chose an $\varepsilon$–strategy or a general strategy. The line for these strategy profiles passed through the interior of the stage game from the region where $S_X > S_Y$ because $q_2+q_3>1$ (or $\varepsilon > 1/2$), to $S_X = S_Y$ because $q_2+q_3=1$ ($\varepsilon = 1/2$), and finally to the region where $S_X < S_Y$ because $q_2+q_3<1$ ($\varepsilon < 1/2$). But $S_X > S_Y$ cannot be considered altruistic from a rational player X point of view because it is the selection of Y strategy set $q_2+q_3>1$ which results in this score ordering. It can be considered altruistic from the point of view of Y because X scores higher due to the strategy set chosen by Y. Similarly, $S_X < S_Y$ cannot be considered extortive from the point of view of the rational X because it is the selection of the Y strategy set $q_2+q_3<1$ which results in this score ordering. This ordering is extortive for Y. The strategic environment is subtle because these score orderings depend on X choosing WSLS. But the WSLS strategy set has a good performance in tournaments [6]. Since the scores $S_X > S_Y$ when X chooses WSLS means that the player Y strategy set with $q_2+q_3>1$ is responsible for good performance of WSLS, one possibility is that player X has prior information that opponents in a tournament are more likely to choose an altruistic strategy set with $q_2+q_3>1$ than to choose the extortive strategy set with $q_2+q_3<1$ [13,22]. Another possibility is that the tournament conditions do not satisfy the conditions for Markov chain analysis. For example, fluctuations may play an important role in the limited repetitions of a human tournament. Fluctuations are beyond the scope of the analysis of the present work. Also Fig. 6 shows a region of no scores around ($P,P$) when WSLS is played. As a final remark, it is important to note that extortive scores for Y over X can exist when player X choses WSLS even without the condition of defection after mutual defection for Y ($q_4 = 0$).

The DaMD strategy has player (X) defecting after previous mutual defection ($p_4 = 0$). However, this property alone is not enough to ensure that X can extort the other player Y [23]. As we show above, the transition probabilities $p_2$ and $p_3$ are also crucial – extortion and reciprocity using the DaMD strategy rest on a three-legged stool with the three legs being the conditional transition probabilities $p_2$, $p_3$, and $p_4$. One leg of course is $p_4 = 0$. The other legs determine whether the scores are extortive ($p_2 + p_3 < 1$ so $S_X > S_Y$) or reciprocal ($p_2 + p_3 = 1$ so $S_X = S_Y$). These results depend on only the DaMD strategy of player X unless player Y also choses a DaMD.

It has been long understood that the Nash equilibrium strategy of a rational player is to defect in a single-play, stage game of Prisoner's Dilemma [13]. A relatively straight-forward argument building on that for the single-play stage game concludes that the equilibrium strategy of a rational player is to defect, certainly on the last play, of a finitely repeated Prisoner's Dilemma game [22]. It has not been clear if such an equilibrium strategy for a rational player exits in the infinitely repeated Prisoner's Dilemma game. The existence of the DaMD strategy provides an answer for the infinitely iterated PD: Two rational players in infinitely repeated Prisoner's Dilemma must conclude that the other rational player choses an extortive DaMD so each player decides to play DaMD with $p_2 + p_3 < 1$ resulting in stationary scores of $(P, P)$, equivalent to the result of mutual defection in the single-play stage game.

It has also long been known that cooperation is a hallmark of human and even some animal behaviors [3,4,6,9,10,11,12]. Given the conclusion above that rational players must essentially play extortive DaMD in the IPD and take the $(P, P)$ payoff, the question of how cooperation arises needs to be considered. This has been considered in the papers on ZDS in evolutionary dynamics already referenced [9,10,11,12,18,20]. Our answer arises within DaMD itself and can lead even rational players out of the $(P, P)$ state. We consider two rational players. Both players know that playing DaMD is a strategy set that can extort a naïve player, but also leads to low mutual scores $(P, P)$ for two rational players. Instead these rational players can choose to initiate the Markov chain using the reciprocal DaMD with $p_1 = 1$, and $p_2 + p_3 = 1$ for player X and the symmetric probability choices for player Y leading to the mutual Pareto optimal scores $(R, R)$. These are the best mutual scores for two rational players in the IPD. If this state is reached then, the two rational players can "go to lunch" and let a random number generator continue play, subject to the conditions of the reciprocal DaMD.

The other situation is that of a rational player vs a naïve player. The rational payer X can extort the naïve player Y by playing an DaMD. But X can be concerned that extortive scores are not stable if Y learns and becomes a rational player and detects that X has a larger score. Y will play the extortive DaMD and, when the Markov stationary state is reestablished, the scores are now $(P, P)$. However, X can choose to play a reciprocal DaMD that results in equal scores and so signal to Y. This is not necessarily altruism on the part of X, but perhaps foresight – Y eventually learns and becomes rational rather than remaining naïve. If the initially naïve Y is playing an $\varepsilon$-strategy, for example, and learns to be rational, then Y can adjust $\varepsilon$ as in Fig. 7 and follows increasing scores to $(R, R)$. Such a path to cooperation has been suggested as a model for international affairs [21]. TFT is a classic DaMD strategy set with reciprocal probabilities since $p_2 + p_3 = 1$ and $S_X = S_Y$. From the point of view of the present work, it is no accident that TFT performed so well in the initial two tournaments of Axelrod – TFT takes the road to cooperation and hence to the $(R, R)$ state. The concepts of altruism [3,4,6], reciprocity [3,4,6], stability [3,6,22], signaling [13,22], and learning [6,22] have deep meaning in game theory. We use the dictionary definitions of these concepts to give a viable description of a path to cooperation that uses reciprocal DaMD as defined herein.

A naïve player always has a score lower than a rational player playing an extortive DaMD unless the naïve player choses an DaMD, perhaps by chance. As a final example, let X select the extortive DaMD given by $\mathbf{p}=(3/4,1/4,1/4,0)$, and Y select an $\varepsilon$-strategy given by $\mathbf{q}=(\varepsilon,\varepsilon,\varepsilon,\varepsilon)$. Using the results in Section VII and choosing Axelrod values, the scores are given by $S_X = (3+10\varepsilon-5\varepsilon^2)/(3-\varepsilon)$ and $S_Y = 3/(3-\varepsilon)$ with extortive ordering $S_X > S_Y$ if $0 < \varepsilon < 1/2$. The equal scores $(1,1)$ occur when $\varepsilon = 0$. To quantify the asymmetry in the extortive scores, we define a distance measure $d(\varepsilon)$ between scores by $d(\varepsilon) = S_X - S_Y = 5\varepsilon(2-\varepsilon)/(3-\varepsilon) \geq 0$. Such a distance measure is related to the concept of frustration that is useful in landscape theory of complexity [19,24,25] and to potential games [22]. In Fig. 8, we plot the scores and the distance measure for rational and naïve players starting at $(1,1)$ so $\varepsilon = 0$. The rational player plays the extortive DaMD above and the naïve player Y randomly increases $\varepsilon$ over 1000 Markov steps to increase Y's own score. Our discussion above suggests that such a relatively large distance of 2.5 and resulting frustration for scores at Markov step 1000 is not be stable if the naïve player learns, becomes rational, and plays an extortive DaMD. We described above a reciprocal strategy profile DaMD that leads to cooperation and the state $(R,R)$. Theoretical explanation of the path to cooperation may ultimately be based on detailed landscape dynamics [19,24,25] and fluctuations [25,26,27]. The present paper provides insight for the development of such a path.

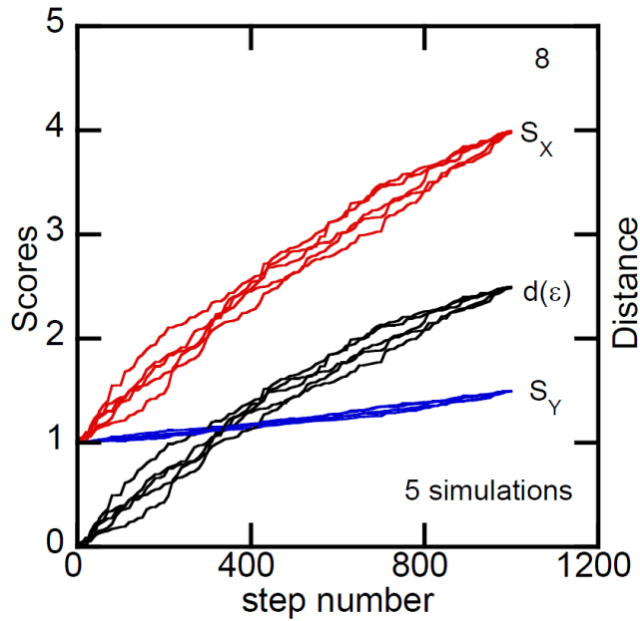

Figure 8. Scores and distance metric vs Markov steps.
Rational extortive DaMD (X) vs Naïve $\varepsilon$-Strategy (Y). See text.

## Acknowledgement

I thank Prof. Stephen G. Walker of Arizona State University for many discussions over the years that eventually interested me enough in game theory to attempt the research reported here.

# References

[1] J. von Neumann and O. Morgenstern, *Theory of Games and Economic Behavior* (Princeton University Press, 2007). Note: 60th Anniversary Commemorative edition.
[2] W. Poundstone, *Prisoner's Dilemma* (Anchor Books, 1993).
[3] R. Axelrod, *The Evolution of Cooperation* (Basic Books, 1984).
[4] K. Sigmund, *The Calculus of Selfishness* (Princeton University Press, 2010).
[5] A. Rapoport, d. A. Seale, A. M. Colman, *PLoS ONE* **10(7)** (2015).
[6] M. A. Nowak, *Evolutionary Games: Exploring the Equations of Life* (Belknap/Harvard University Press, 2006).
[7] G. Lambert, S. Vyawahare, and R. H. Austin, *Interface Focus* **4**, 20140029 (2014).
[8] W. H. Press and F. J. Dyson, *Proc. Nat. Acad. Sci. USA* **109**, 10409 (2012).
[9] C. Hilbe, M. A. Nowak, and K. Sigmund, *Proc. Nat. Acad. Sci. USA* **110**, 6913, (2013).
[10] Z. Wu and Z. Rong, *Phys. Rev. E* **90**, 062102 (2014).
[11] C. Hilbe, B. Wu, A. Traulsen, and M. A. Nowak, *J. Theor. Biol.* **374**, 315 (2015).
[12] X. Xu, Z. Rong, Z. Wu, T. Zhou, and C. K. Tse, *Phys. Rev. E* **95**, 052302 (2017).
[13] S. Tadelis, *Game Theory: An Introduction* (Princeton University Press, 2013).
[14] K. Leyton-Brown and Y. Shoham, *Essentials of Game Theory: A Concise, Multidisciplinary Introduction* (Morgan and Claypool, 2008).
[15] J. Nash, *Proc. Nat. Acad. Sci. USA* **36**, 48 (1950).
[16] S. Karlin and H. M. Taylor, *A First Course in Stochastic Processes, 2nd ed.* (Academic Press, 1975).
[17] N. Privault, *Understanding Markov Chains: Examples and Applications* (Springer, 2013).
[18] C. Hilbe, T. Röhr, and M. Milinski, *Nat. Comm.* **5**, 3976 (2014).
[19] L. Xu and J. Wang, PLoS One **13**, e0201130 (2018).
[20] Z. Wang, Y. Zhou, J. W. Lien, J. Zheng, and B. Xu, *Nat. Comm.* **7**, 11125 (2016).
[21] R. Liao, *Georgetown J. Int. Affairs* **17**, 38 (2016).
[22] Y. Shoham and K. Leyton-Brown, *Multiagent Systems: Algorithmic, Game-Theoretic, and Logical Foundations* (Cambridge University Press, 2009).
[23] S. Wang and F. Lin, arXiv:1712.06488.
[24] R. Axelrod and D. S. Bennett, *Brit. J. Pol. Sci.* **23**, 211 (1993).
[25] P. W. Fenimore, H. Frauenfelder, B. H. McMahon, and R. D. Young, *Proc. Nat. Acad. Sci. USA* **101**, 14408 (2004).
[26] D. K. C. MacDonald, *Noise and Fluctuations*, (Dover, 2006).
[27] S. Miller and J. Knowles, *Sci. Rep.* **5**, 11054 (2015).